\documentclass[aps,prl,twocolumn,groupedaddress]{revtex4}

\begin{document}

%\preprint{}
\title{Critical Tests for two Hypotheses used in Gravitation and Cosmology}


\author{Rafael A. Vera}
\email[rvera@udec.cl]
%\homepage[http://www2.udec.cl/\symbol{126}rvera]
%\thanks{}
\affiliation{Facultad de Ciencias F\'{i}sicas y Matem\'{a}ticas\\
Universidad de Concepci\'{o}n. Chile}


\date{\today}

\begin{abstract}
From more critical tests for gravitational (G) hypotheses it has
been proved that the relative properties of nonlocal (NL) bodies at
rest with respect to an observer depend on the difference of G
potential between bodies and observer. The G energy comes not from
the field but from a fraction of the mass-energy of the bodies.
These results are in opposition with two traditional hypotheses used
in Physics. In the classical tests of general relativity their
errors are compensated because they have the same absolute value and
opposite signs. The general theory based on such tests is consistent
with quantum mechanics and with all of the traditional G tests. The
new cosmological scenario, radically different from the standard
one, has also been verified from astronomical observations.
\end{abstract}

%\keywords{}

\maketitle

\section{\label{1}Introduction}

To test gravitational (G) hypotheses, a general theory not based on
any of them, called ``\textit{nonlocal (NL) relativity}'' (NLR), was
introduced by R. Vera in the Einstein's Centennial Symposium on
Fundamental Physics (Bogot\'{a}, 1979) [1] and in a later paper [2].
\textit{It is based on the Einstein's equivalence principle} (EEP)
according to which ``radiation in stationary state between two well
defined parts of a system must have the same inertial and G
properties of the other parts of the same system''. Thus \emph{the
properties of a particle model }(PM)\emph{made up of a photon in
stationary state, derived from dual properties of light, are
consistent with special relativity, quantum mechanics and all of the
conventional G test}[1,2,3]

During a free fall, the acceleration of a PM is due to a plain
refraction phenomenon in which \emph{its relative
mass-energy-frequency with respect to an observer at rest is
conserved}, which is \emph{opposed to the Einstein's G field energy
hypothesis} (GFEH). \emph{The energy released comes from a fraction
of its mass-energy lost during the stop}. Thus the final relative
mass-energy of the PM at rest with respect to the observer depends
on the difference of GP between the body and the observer, which is
\emph{opposed to the classical hypothesis }(CH) in that the relative
rest mass of the NL body with respect to the observer is identical
to the local one.

Below, these results have been verified from other critical tests
that are independent on G hypotheses.

\section{\label{2}A critical test from optical physics}

Assume that a standing electromagnetic wave exists between two
mirrors at rest in the positions $A$ and $B$ of a central field.
From ``wave continuity'', the number ($N$)of waves between $A$ and $B$%
, and the relative time interval of the trip $AB$ of ``each wave''
with respect to the clock at $A$, called $\Delta t_{A}(B),$ are
strictly constants. Thus \emph{the relative frequency of the waves
reflected at $A$\ and $B$, with respect to the clock at $A$, is the
same}:

\begin{equation}
\nu _{A}(A)=\nu _{A}(B)=\frac{N_{AB}}{\Delta t_{A}(B)}= Constant_{A}
\label{1}
\end{equation}

Then \textit{the relative frequencies and energies of photons
traveling freely in a G field, with respect to an observer at rest
in the field, remain constants, respectively, i.e., G fields don't
exchange energy with photons which is in clear disagreement with the
GFEH.}

From (1) the relative ``emission frequencies'' of a NL atom at rest
at $B$ with respect to the observer at $A$, called $\nu _{A}(0,B)$,
are just equal to the ones measured locally in G red shift
experiments (GRSE).

\begin{equation}
\nu _{A}(0,B)=\nu _{A}(B)=\nu _{A}(A)  \label{2}
\end{equation}

Thus the result of a GRSE can be described by:

\begin{equation}
\frac{\Delta \nu _{A}(0,B)}{\nu _{A}(0,A)}=\Delta \phi _{A}(B)=\frac{%
\Delta E_{A}(0,B)}{m_{A}(0,A)}  \label{3}
\end{equation}
in which $\Delta \nu _{A}(0,B)=\nu _{A}(0,B)-\nu _{A}(0,A)$.

The proportional difference of the emission frequency of the NL
atoms at B with respect to the observer at A is equal to the
difference of GP defined by the last member in terms of the
proportion of mass-energy released by the atoms, or a test body,
after a free fall from $B$ and stop at $A$. Such frequency
differences existed before the photons were emitted, which is in
opposition with the CH.

\section{\label{3}A critical test from G time dilation}

Here, a G time dilation experiment (GTDE) is the result of comparing
well-defined time intervals of local and nonlocal clocks at rest in
different GP. \emph{This is a critical test because it is
independent on the frequency of any radiation traveling between the
clocks, i.e, it is independent on results and hypotheses used in
GRSE}.

The results of GTDE are identical to those given by (3). The
proportional frequency differences of clocks and atoms are the same.

On the other hand, from the EEP, the frequencies, masses and lengths
of each part of a system are related to each other, locally, by
universal constants:

\begin{equation}
\nu _{A}(0,A):\lambda _{A}(0,A):m_{A}(0,A):=C1:C2:C3  \label{4}
\end{equation}

If a body changes of rest position from $A$ to $B=A+\Delta r$ and
the observer stays at $A$, from (3) and (4),

\begin{equation}
\frac{\Delta \nu _{A}(0,B)}{\nu _{A}(0,A)}=\frac{\Delta \lambda _{A}(0,B)}{%
\lambda _{A}(0,A)}=\frac{\Delta m_{A}(0,B)}{m_{A}(0,A)}=\frac{\Delta E_{A}(B)%
}{m_{A}(0,A)}  \label{5}
\end{equation}

\textit{The proportional differences of the relative properties of a
NL body at rest with respect to the observer, are just equal to the
differences of GP between the body and the observer. } This result
may be called \textit{the non-equivalence principle} (NEP) \emph{for
bodies at rest in different} GP.

From (5) and special relativity applied to a free fall from $B$ and
a stop at $A$ it is inferred that:

\begin{equation}
m_{A}(0,B)=m_{A}(V,A)=m_{A}(0,A)+\Delta E_{A}(B)  \label{6}
\end{equation}

\textit{During a free fall, the relative mass-energy of the NL body
with respect to the observer at rest at} $A$ \emph{remains constant.
This value is higher than the local one at rest at}$A$. \emph{Thus
both the CH and the GFEH would be wrong}.

\section{\label{4}Astronomical tests}

From (5), a uniform universe expansion cannot produce measurable
changes with the time because the increase of GP would expand the
reference standards in identical proportion, regardless on their
internal structures. Thus the cosmological red-shifts don't change
during expansion, i.e. \textit{the universe age may be infinite.}

From the lack of energy of the G field, the new kind linear black
hole (LBH) would be a giant nucleus around which the high gradient
of the relative refraction index would prevent, by critical
reflection, the escape of radiation and relativistic particles.
After absorbing radiation, for a long period, the relative
mass-energy of its nucleons can be equal or larger than the mass of
a free neutron far from the LBH. In such ``excited state'', the LBH
can decay into a cloud of primeval gas thus starting a new evolution
cycle of matter.

Then galaxies should be evolving, rather indefinitely, in nearly
closed cycles with luminous and dark periods. The LBHs formed during
the luminous periods, after a long energy absorption period, would
explode, in chains, generating gas for a luminous period and so
on[1,2,3,4,5].

It may be verified that all of the different evolution phases of a
galaxy cycle are present anywhere in the universe. They are
consistent both with the different kinds of galaxies, luminous,
partially luminous, and dark ones, found in the local universe and
in the deep field observations. Statistically, most of them should
be in their dark states, absorbing energy from the rest of the
universe. The last ones should account for the dark matter in the
universe and the low temperature CMB [5,6,7,8].

\section{\label{5}Conclusions}

From NLR and tests more critical than the classical ones it is found
that two current G hypotheses would be wrong. This error is not easy
to detect because such hypotheses are often used altogether, like in
the classical G tests. In such way their errors are canceled out.

After using a particle model consistent with the EEP, instead of G
hypotheses, NLR accounts for a wider range of inertial and
gravitational properties of uncharged bodies, including special
relativity, quantum mechanics, the classical G tests and more
critical ones. From them, the universe has no limits of age for the
evolution of galaxies in nearly closed cycles, which can be verified
from the observations of their different phases anywhere in the
universe. Thus the dark matter and the low temperature CMB turn out
to be critical verifications of NLR.

\end{document}